\documentclass{article}
\usepackage{amssymb}
\usepackage{amsmath}
\usepackage{graphicx}

\begin{document}

\title{The Gross-Pitaevskii-Poisson model for an ultracold plasma: density
waves and solitons}
\author{Hidetsugu Sakaguchi$^{1}$ and Boris A. Malomed$^{2,3}$ \\
$^{1}$Department of Applied Science for Electronics and Materials,\\
Interdisciplinary Graduate School of Engineering Sciences,\\
Kyushu University, Kasuga, Fukuoka 816-8580, Japan\\
$^{2}$Department of Physical Electronics, School of Electrical Engineering,\\
Faculty of Engineering, and Center for Light-Matter Interaction,\\
Tel Aviv University, Tel Aviv 69978, Israel\\
$^3$Instituto de Alta Investigaci\'{o}n, Universidad de Tarapac\'{a}, Casilla 7D, Arica,
Chile}
\maketitle

\begin{abstract}
We introduce one- and two-dimensional (1D and 2D) models of a degenerate
bosonic gas composed of ions carrying positive and negative charges (cations
and anions), under the condition of the electro-neutrality. The system may
exist in the mean-field condensate state, enabling the competition of the
Coulomb coupling, contact repulsion, and kinetic energy of the particles,
provided that their effective mass is reduced by means of a lattice
potential. The respective model combines the Gross-Pitaevskii (GP)\
equations for the two-component wave function of the cations and anions,
coupled to the Poisson equation for the electrostatic potential mediating
the Coulomb coupling. In addition to its direct introduction, the contact
interaction in the GP system can be derived, in the Thomas-Fermi
approximation, from a system of three GP equations, which includes the wave
function of heavy neutral (buffer) atoms. In the system with fully repulsive
contact interactions, we construct stable spatially periodic patterns
(density waves, akin to ionic crystals). The transition to the density wave is identified by
analysis of the modulational instability of a uniformly mixed neutral state.
The density-wave pattern, which represents the system's ground state (GS),
is accurately predicted by a variational approximation. In the 2D case, a
stable pattern is produced too, with a quasi-1D shape. The 1D system with
contact self-attraction in each component produces bright solitons of three
types: neutral ones, with fully mixed components; dipoles, with the
components separated by the inter-species contact repulsion; and
quadrupoles, with a layer of one component sandwiched between side lobes
formed by the other. The transition from the neutral solitons to dipoles is
accurately modeled analytically. A chart of the GSs of the different types
(neutral solitons, dipoles, or quadrupoles) is produced. Different soliton
species do not coexist as stable states.Collisions between traveling
solitons are studied too. Collisions are elastic for dipole-dipole pairs,
while dipole-antidipole ones merge into stable quadrupoles via multiple
collisions.
\end{abstract}

\section{Introduction and the model}

For Bose-Einstein condensates (BECs), created in ultracold atomic gases \cite%
{Anderson}, an exceptionally accurate dynamical model is provided by the
Gross-Pitaevskii (GP) equation \cite{GP}. The GP equation was derived for
BEC with contact inter-particle interactions, and extended for \ the gas of
dipolar atoms, with long-range interactions between them. Various aspects of
theoretical and experimental studies of dipolar condensates are summarized
in Refs. \cite{dipole-review1}-\cite{dipole-review3}. Further, the analysis
was developed, chiefly in the context of astrophysical and cosmological
models, for BEC bound by gravity forces \cite{gravity1}-\cite{Yukalovs}. It
was also demonstrated that similar long-range attractive interactions can be
artificially induced in atomic condensates by means of specially designed
laser illumination \cite{Kurizki,gravity emulation}. A natural model for the
BEC with gravitational or pseudo-gravitational interactions is a system of
the GP equation for the mean-field atomic wave function and Poisson equation
for the gravitational potential. A model of BEC in the gas of particles
carrying electric dipole moments is also based on the GP-Poisson (GPP)
system, making use of two different mean-field approximations: one for the
particles' wave function, and another for the interaction of the particle's
dipole moment with the electrostatic potential, which is created, as per the
Poisson equation, by the distribution of the polarization density in the gas
\cite{we}.

The current work with ultracold plasmas \cite{Killian-PRL}-\cite{Rolston}
suggests a possibility to consider them in the state of quantum degeneracy
\cite{Manfredi1}-\cite{Fedele}. Although presently available experimental
techniques make it possible to cool ions down to the lowest level of $\sim
10~\mathrm{\mu }$K \cite{Killian-PRL}, \cite{NJP}-\cite{Holland}, which
remains ca. two orders of magnitude above the temperature of the BEC
transition, the ongoing progress in the experimental studies makes is
plausible that a plasma in the BEC state will be eventually created.

The theoretical analysis of quantum plasmas was developed for the
electron-ion mixtures \cite{Manfredi1,Manfredi2,Shukla}. They are modeled by
the GPP system, including the nonlinear Schr\"{o}dinger (NLS)\ equation for
the wave function of electrons, $\Psi $, coupled to the electrostatic
potential. In turn, the potential is created by the respective density, $%
\left\vert \Psi \right\vert ^{2}$, pursuant to the Poisson equation. The NLS
equation for $\Psi $ includes a self-repulsion term $\sim \left\vert \Psi
\right\vert ^{4/D}\Psi $, where $D$ is the spatial dimension; thus, it is
similar to the effective quintic nonlinearity of the Tonks-Girardeau gas in
the 1D setting \cite{Kolomeisky}, usual cubic nonlinearity of BEC\ \cite{GP}
in 2D, and the nonlinear term known in the density-functional model of Fermi
gases \cite{Fermi1,Fermi2}, which may be applied to experimentally available
settings \cite{Jin}.

A less common type of plasmas represents mixture of positive and negative
ions (cations and anions), containing a negligibly small density of free
electrons. Such plasmas were created with different \cite%
{diff-mass1,diff-mass2} and equal \cite{equal-mass} cation and anion masses.
Theoretically, an ultracold plasma composed of cations and anions with equal
masses was addressed in Ref. \cite{positive-negative}, by means of the
molecular-dynamics methods, i.e., numerically solving coupled equations of
motion for all particles in the plasma. In particular, known results \cite%
{deuterium} suggest that a candidate for achieving this purpose may be a
mixture of deuterium anions and cations.

Assuming that an electroneutral cation-anion plasma may be, eventually,
cooled down into the BEC state, we here introduce a model for it, which
includes a two-component GP equation for mean-field wave functions $\psi
_{\pm }$ of the positive and negative bosonic ions, and the Poisson equation
for the electrostatic potential, $\phi $, induced by the distribution of the
local charge density. The objective is to produce stable GPP states in the
form of spatially-periodic density waves and bright solitons. Chiefly, the
results are reported for the effectively 1D setting, and some findings for
density waves are obtained for the 2D system too.

Before introducing the model written in a scaled form, it is relevant to
estimate possible physical parameters of the setting under the
consideration. The state of degenerate plasmas is controlled by the
Coulomb-coupling strength $C$, \textit{viz}., ratio of the energy of the
electrostatic interaction between charge carriers to their kinetic energy
 \cite{Ichimaru} (the latter energy term is restricted by the condition of degeneracy
of the gas). For a high-density plasma (up to $10^{18}$ cm$^{-3}$) of light cations
and anions, an estimate yields $C\sim 10^{4}$ (very strong Coulomb interaction is
actually a reason impeding the creation of quantum plasmas in the experiment).
At $C\gtrsim 10^{3}$, the strongly-coupled plasma is expected to
build states such as Wigner crystals \cite{Astra-Girar}, different from
those predicted by the GP theory, as the gradient terms becomes negligible
in it. The plasma may be kept in the regime of moderate Coulomb coupling if
the effective ion mass, $m^{\ast }$, is made essentially smaller than its
bare value, by means of a lattice (spatially periodic) potential. Lattice
potentials were used for such purposes in many theoretical and experimental
studies, making use of the fact that the effective mass may be strongly
reduced close to edges of the bandgap in the respective spectrum \cite%
{Pit,Smerzi}. In particular, it was demonstrated theoretically \cite%
{Carusotto} and experimentally \cite{Inguscio} that one may easily reduce $%
m^{\ast }$ by up to two order of magnitude, against the bare value.

As concerns the comparison of the Coulomb and kinetic energies, it is
relevant to stress that the patterns reported below are actually produced by
the competition of the electrostatic interaction not with the quantum
dispersion (which corresponds to the kinetic energy in the GP equation),
but, chiefly, with contact interactions between the ions (solutions for the
density waves and solitons do not exist in the absence of the contact
interactions). For the same typical values of physical parameters which
produce the above-mentioned estimate, $C\sim 10^{4}$, the ratio of the
contact-interaction and kinetic energies is estimated as being $\sim
10^{2}-10^{3}$, if the the bare atomic mass is used. Thus, the introduction
of the reduced effective mass may make magnitudes of the latter energy terms
comparable, which helps to explain the possibility of the creation of the
spatial patterns considered below.

According to what is said above, the GPP system is written, in the scaled
form, as%
\begin{eqnarray}
i\frac{\partial \psi _{+}}{\partial t} &=&\left[ -\frac{1}{2m_{+}^{\ast }}%
\nabla ^{2}+\left( g_{++}\left\vert \psi _{+}\right\vert
^{2}+g_{+-}\left\vert \psi _{-}\right\vert ^{2}\right) +q_{+}\phi
+U_{+}\left( \mathbf{r}\right) \right] \psi _{+},  \label{++} \\
i\frac{\partial \psi _{-}}{\partial t} &=&\left[ -\frac{1}{2m_{-}^{\ast }}%
\nabla ^{2}+\left( g_{--}\left\vert \psi _{-}\right\vert
^{2}+g_{+-}\left\vert \psi _{+}\right\vert ^{2}\right) -q_{-}\phi
+U_{-}\left( \mathbf{r}\right) \right] \psi _{-},  \label{--} \\
\nabla ^{2}\phi &=&-4\pi \left( q_{+}\left\vert \psi _{+}\right\vert
^{2}-q_{-}\left\vert \psi _{-}\right\vert ^{2}\right) ,  \label{phiphi}
\end{eqnarray}%
where $m_{\pm }^{\ast }$ and $q_{\pm }$ are scaled effective masses and
absolute values of charges of the cations and anions, $g_{++,--,+-}$ are
coefficients of the contact nonlinearity, and $U_{\pm }\left( \mathbf{r}%
\right) $ are trapping potentials, if any (if the lattice potential is used,
as outlined above, to renormalize the effective masses, it is not explicitly
included, as its effect is represented by $m_{\pm }^{\ast }$). Equations (%
\ref{+})-(\ref{phi}) conserve two numbers of ions,
\begin{equation}
N_{\pm }=\int \left\vert \psi _{\pm }\left( \mathbf{r}\right) \right\vert
^{2}d\mathbf{r},  \label{N}
\end{equation}%
where $d\mathbf{r}=dx$, $dxdy$, or $dxdydz$, in the 1D, 2D, and 3D cases,
respectively, and the electroneutrality condition implies%
\begin{equation}
q_{+}N_{+}=q_{-}N_{-}~.  \label{neutral}
\end{equation}

Below, we focus on the consideration of the symmetric system in free space ($%
U_{\pm }=0$), with equal effective masses which are fixed, by rescaling the
coordinates, to be $m_{+}^{\ast }=m_{-}^{\ast }\equiv 1$; further, the
symmetry implies $g_{++}=g_{--}\equiv g$, $g_{+-}\equiv G$, and $%
q_{+}=q_{-}\equiv q$. Then, $q_{+}=q_{-}$ implies $N_{+}=N_{-}\equiv N$, as
per Eq. (\ref{neutral}). Finally, making use of the remaining scale
invariance of the GPP system, we set $g=\pm 1$ for the repulsive and
attractive signs of the contact self-interaction, respectively, unless $g=0$%
. Thus, the symmetric version of the GPP system of Eqs. (\ref{++})-(\ref%
{phiphi}) takes the form of%
\begin{gather}
i\frac{\partial \psi _{+}}{\partial t}=\left[ -\frac{1}{2}\nabla ^{2}+\left(
g\left\vert \psi _{+}\right\vert ^{2}+G\left\vert \psi _{-}\right\vert
^{2}\right) +q\phi \right] \psi _{+},  \label{+} \\
i\frac{\partial \psi _{-}}{\partial t}=\left[ -\frac{1}{2}\nabla ^{2}+\left(
g\left\vert \psi _{-}\right\vert ^{2}+G\left\vert \psi _{+}\right\vert
^{2}\right) -q\phi \right] \psi _{-},  \label{-} \\
\nabla ^{2}\phi =-4\pi q\left( \left\vert \psi _{+}\right\vert
^{2}-\left\vert \psi _{-}\right\vert ^{2}\right) ,  \label{phi}
\end{gather}

In fact, the same system can be derived in the framework of a more general
setting, which includes, in addition to the cations and anions, a buffer
component of heavy neutral bosonic atoms with wave function $\Psi _{0}$ (the
neutral atoms may be also used for cooling the ions by means of the
sympathetic method \cite{Weizmann2,Holland}). Thus, in the limit case when
the contact interactions are much weaker than the electrostatic coupling
mediated by potential $\phi $ (which is a natural situation, as mentioned
above), two GP equations (\ref{+}) and (\ref{-}) are replaced by three, the
extra one being the equation for $\Psi _{0}$:
\begin{gather}
i\frac{\partial \psi _{+}}{\partial t}=\left[ -\frac{1}{2}\nabla
^{2}+f\left\vert \Psi _{0}\right\vert ^{2}+q\phi \right] \psi _{+},
\label{2+} \\
i\frac{\partial \psi _{-}}{\partial t}=\left[ -\frac{1}{2}\nabla
^{2}+f\left\vert \Psi _{0}\right\vert ^{2}-q\phi \right] \psi _{-},
\label{2-} \\
i\frac{\partial \Psi _{0}}{\partial t}=\left[ -\frac{1}{2M}\nabla
^{2}+F\left\vert \Psi _{0}\right\vert ^{2}+f\left( \left\vert \psi
_{+}\right\vert ^{2}+\left\vert \psi _{-}\right\vert ^{2}\right) \right]
\Psi _{0},  \label{20}
\end{gather}%
where $M$ is the relative mass of the buffer atoms, while $f$ and $F$ are
coefficients accounting for, respectively, the buffer-ion and self-buffer
interactions. For heavy atoms with large $M$, the kinetic-energy term in Eq.
(\ref{20}) may be neglected, which is tantamount to the Thomas-Fermi
approximation \cite{GP}, applied to this equation. Then, looking for a
solution as $\Psi _{0}=\exp \left( -i\mu _{0}t\right) \psi _{0}\left(
x,y,z;t\right) $ and $\psi _{\pm }\equiv \exp \left( -i(f/F)\mu _{0}t\right)
\tilde{\psi}_{\pm }$, where $\mu _{0}$ is the buffer's chemical potential,
and $\psi _{0}$ is a slowly varying function of $t$ in comparison with $\exp
(-i\mu _{0}t)$, one can use Eq. (\ref{20}) to eliminate the buffer's
density,
\begin{equation}
\left\vert \psi _{0}\right\vert ^{2}=F^{-1}\left[ \mu _{0}-f\left(
\left\vert \tilde{\psi}_{+}\right\vert ^{2}+\left\vert \tilde{\psi}%
_{-}\right\vert ^{2}\right) \right] .  \label{psi0}
\end{equation}%
The substitution of this approximation in Eqs. (\ref{2+}) and (\ref{2-})
leads back to Eqs. (\ref{+}) and (\ref{-}) for wave functions $\tilde{\psi}%
_{\pm }$, with effective coefficients
\begin{equation}
\tilde{g}=\tilde{G}=-f^{2}/F.  \label{tilde}
\end{equation}

Spatial patterns are expected to exist if the Coulomb coupling between
cations and anions is balanced by the contact repulsion between them,
therefore we adopt $G>0$ in Eqs. (\ref{+}) and (\ref{-}) [in particular,
this implies the choice of $F<0$, i.e., self-attraction of the buffer atoms,
if the nonlinearity coefficients are produced by Eq. (\ref{tilde})].

It may be interesting to consider the system which includes, on a par with
the cations and anions, polar molecules formed as their bound states, which
should be represented by a separate wave function. The respective model
should be based on a system of three GP equations, including various
\textquotedblleft reactions", such as merger of colliding ions into the
molecule [however, the merger may be suppressed by the strong contact
repulsion with $G>0$ in Eqs. (\ref{+}) and (\ref{-})], and breakup of the
molecule due to collisions. In the present paper, we do not aim to address
such a system.

Note also that Eqs. (\ref{+}) and (\ref{-}) with $g=0$ [or with $g$ given
by Eq. (\ref{tilde}), if the ion-ion interaction is mediated by the buffer
atoms] may model a fermionic plasma \cite{Fermi}, in which the Pauli
principle forbids direct self-interaction, cf. Ref. \cite{mediated}.
Furthermore, the derivation of the NLS-Poisson system for the quantum plasma
dominated by the electron component \cite{Manfredi1,Manfredi2,Shukla}
suggests that, in the case of $g=+1$ and imaginary $G=-i\Gamma $, the 2D
version of Eqs. (\ref{+})-(\ref{phi}) may serve as a model of the degenerate
electron-positron plasma, with coefficient $\Gamma >0$ representing
annihilation losses, although the consideration of this possibility is
beyond the scope of the present work.

The rest of the paper is organized as follows. Spatially-periodic
density-wave solutions, in both 1D and 2D forms, are considered below in
Section II. That section includes an exact analytical investigation of the
modulational instability (MI) of the uniformly mixed (locally neutral)
state, and an analytical approximation for the density-wave patterns.
Solutions for 1D solitons, which may exist in the form of neutral localized
states, with fully mixed cations and anions, or dipole and quadrupole ones,
are addressed in Section III. In particular, the transition from neutral
solitons to dipoles is predicted analytically. The chart of the system's
ground states (GSs), represented by the neutral, dipole, or quadrupole
solitons, is produced in a numerical form. Collisions between moving dipole
solitons are considered too, by means of direct simulations. The paper is
concluded by Section IV.

\section{Spatially periodic states}

\subsection{Stationary equations}

Stationary 1D solutions of Eqs. (\ref{+})-(\ref{phi}) with chemical
potentials $\mu _{\pm }$ of the two components are looked for as
\begin{equation}
\psi _{\pm }\left( x,t\right) =e^{-i\mu _{\pm }t}u_{\pm }(x),  \label{u}
\end{equation}%
with real wave functions $u_{\pm }(x)$ obeying the stationary version of the
GPP system in free space ($U_{\pm }=0$):%
\begin{gather}
\left( \mu _{+}-q\phi \right) u_{+}=-\frac{1}{2}\frac{d^{2}u_{+}}{dx^{2}}%
+\left( gu_{+}^{2}+Gu_{-}^{2}\right) u_{+},  \label{+u} \\
\left( \mu _{-}+q\phi \right) u_{-}=-\frac{1}{2}\frac{d^{2}u_{-}}{dx^{2}}%
+\left( gu_{-}^{2}+Gu_{+}^{2}\right) u_{-},  \label{-u} \\
\frac{d^{2}\phi }{dx^{2}}+4\pi q\left( u_{+}^{2}-u_{-}^{2}\right) =0.
\label{phi2}
\end{gather}%
Note that Poisson equation (\ref{phi}) can be solved by means of its Green's
function, which yields
\begin{equation}
\phi (x)=-2\pi q\int_{-\infty }^{+\infty }\left\vert x-x^{\prime
}\right\vert \left[ \left( u_{+}(x^{\prime })\right) ^{2}-\left(
u_{-}(x^{\prime })\right) ^{2}\right] dx^{\prime }  \label{Green}
\end{equation}%
[cf. Ref. \cite{Dong}, where the Green's function was used to solve the
Poisson equation for microwave field coupling two different states of
neutral atoms, represented by two components of a spinor wave function].
Accordingly, Eqs. (\ref{+u}) and (\ref{-u}) may be replaced by a system of
integrodifferential equations:%
\begin{gather}
\mu _{+}u_{+}=-\frac{1}{2}\frac{d^{2}u_{+}}{dx^{2}}+\left(
gu_{+}^{2}+Gu_{-}^{2}\right) u_{+}-2\pi qu_{+}\int_{-\infty }^{+\infty
}\left\vert x-x^{\prime }\right\vert \left[ \left( u_{+}(x^{\prime })\right)
^{2}-\left( u_{-}(x^{\prime })\right) ^{2}\right] dx^{\prime },
\label{integro+} \\
\mu _{-}u_{-}=-\frac{1}{2}\frac{d^{2}u_{-}}{dx^{2}}+\left(
gu_{-}^{2}+Gu_{+}^{2}\right) u_{-}+2\pi qu_{-}\int_{-\infty }^{+\infty
}\left\vert x-x^{\prime }\right\vert \left[ \left( u_{+}(x^{\prime })\right)
^{2}-\left( u_{-}(x^{\prime })\right) ^{2}\right] dx^{\prime }.
\label{integro-}
\end{gather}

Strictly speaking, the Poisson equation should be taken in the 3D form, even
if the atomic components are confined to the effectively 1D setting, as
usual, with the help of a tight confining potential applied in the
perpendicular plane \cite{Salasnich,Delgado}. However, the logarithmic form
of the fundamental solution of the 2D Poisson equation in the orthogonal
plane implies that the transverse component of the electric field will give
rise to a relatively weak force acting on the ions, and their transverse
motion will be suppressed by the confining potential.

\subsection{Modulational instability (MI)\ of the uniformly mixed neutral
state}

First, the GPP system of Eqs. (\ref{+u})-(\ref{phi2}) gives rise to obvious
uniformly mixed (locally neutral) states, with
\begin{equation}
\phi =0,~u_{\pm }=\mathrm{const}\equiv u_{0},~\mu _{\pm }=\left( g+G\right)
u_{0}^{2}.  \label{uniform}
\end{equation}%
These states may be subject to MI\ against perturbations demixing the two
components. The well-known condition for the instability of the mixed state
against the phase separation in the absence of the Coulomb interaction is $%
G>g$ \cite{Mineev}). In the present case, one may expect that the
instability threshold is shifted to larger values of $G$, as the Coulomb
attraction between the components tends to enhance the trend to their mixing.

To investigate the MI, we follow the usual approach, substituting, in the 1D
version of GP equations (\ref{+})-(\ref{phi}), the amplitude-phase form of
the wave functions, $\psi _{\pm }\left( x,t\right) =u_{\pm }\left(
x,t\right) \exp \left( i\chi _{\pm }\left( x,t\right) \right) $ \cite%
{Agrawal}. Then, equations are linearized for modulational perturbations, $%
\delta u_{\pm }=u_{\pm }-u_{0}$, $\chi _{\pm }$, and $\phi $. Solutions for
eigenmodes of the small perturbations are looked for as
\begin{equation}
\left( \delta u_{\pm },\chi _{\pm },\phi \right) =\left( \delta u_{\pm
}^{(0)},\chi _{\pm }^{(0)},\phi ^{(0)}\right) \exp \left( \gamma
t+ipx\right) ,  \label{pert}
\end{equation}%
where $\left( \delta u_{\pm }^{(0)},\chi _{\pm }^{(0)},\phi ^{(0)}\right) $
and $p$ are amplitudes and an arbitrary wavenumber of the perturbation,
while $\gamma $ is the respective MI gain.

The substitution of perturbations (\ref{pert}) in the linearized GPP system
leads to the dispersion equation, relating $\gamma $ and $p$, which is
written in the form of the corresponding determinant in Appendix. Finally, a
straightforward calculation yields four branches of $\gamma (p)$:
\begin{gather}
\gamma =\pm ip\sqrt{\left( g+G\right) u_{0}^{2}+p^{2}/4},  \label{g1} \\
\gamma =\pm \sqrt{Gu_{0}^{2}p^{2}-u_{0}^{2}\left( 8\pi q^{2}+gp^{2}\right)
-p^{4}/4}.  \label{g2}
\end{gather}%
In this section, we consider the case of repulsive self-interactions, with $%
g>0$. Then, the branch of the dispersion relation given by Eq. (\ref{g1}) is
purely imaginary and does not lead to MI. On the other hand, branch (\ref{g2}%
) produces real values of $\gamma $, which represent MI, when the density of
the uniform neutral state exceeds a critical value (i.e., the nonlinearity
is strong enough in comparison with the Coulomb attraction):%
\begin{equation}
u_{0}^{2}>\left( u_{0}^{2}\right) _{\mathrm{cr}}=\frac{8\pi q^{2}}{G-g}.
\label{cr}
\end{equation}%
At $u_{0}^{2}=\left( u_{0}^{2}\right) _{\mathrm{cr}}$, the MI emerges at
wavenumbers with%
\begin{equation}
p^{2}=\left( p^{2}\right) _{\mathrm{cr}}\equiv 16\pi q^{2}.  \label{pcr}
\end{equation}%
The largest value of the squared MI gain,%
\begin{equation}
\gamma _{\max }^{2}=u_{0}^{2}\left[ \left( G-g\right) u_{0}^{2}-8\pi q^{2}%
\right] ,  \label{max}
\end{equation}%
is attained at $p_{\max }^{2}=2\left( G-g\right) u_{0}^{2}$, which is
tantamount to the value given by Eq. (\ref{pcr}) at the MI-onset point, $%
u_{0}^{2}=\left( u_{0}^{2}\right) _{\mathrm{cr}}$. Note that $p_{\max }$
does not depend on the charge, $q$, being the same as in the case of the MI
against demixing of the two components in the absence of the Coulomb
interactions \cite{Mineev}. On the other hand, the critical value of the
squared wavenumber, given by Eq. (\ref{pcr}), does not depend on
coefficients $G$ and $g$ of the contact interactions.

\begin{figure}[h]
\begin{center}
\includegraphics[height=5.0cm]{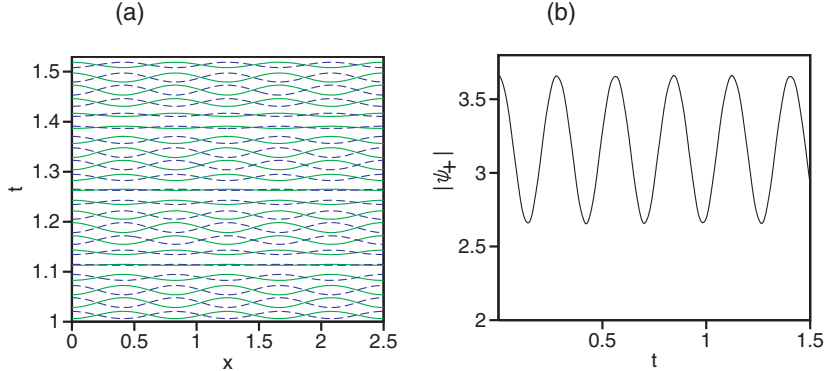}
\end{center}
\caption{(a) Snapshot profiles of $|\protect\psi _{+}(x,t)|$ and $|\protect%
\psi _{-}(x,t)|$ (green solid and blue dashed lines, respectively), at $%
t=0.025n$, with $n=40,41,\cdots ,61$, produced by simulations of Eqs. (%
\protect\ref{+})-( \protect\ref{phi}) with $G=2$ and $g=q=1$, in the domain
of size $L=10$ with periodic boundary conditions, starting from input (%
\protect\ref{input}). (b) The evolution of $|\protect\psi _{+}|$ at $x=L/2$.
}
\label{f0}
\end{figure}

The predicted stability of the uniform neutral state against modulational
perturbations at $u_{0}^{2}\leq \left( u_{0}^{2}\right) _{\mathrm{cr}}$ is
readily confirmed by direct simulations of Eqs. (\ref{+})-(\ref{phi}) in 1D.
As an example, Fig. \ref{f0}(a) displays the evolution initiated by input
\begin{equation}
\psi _{\pm }(x,t=0)=\sqrt{10}\pm 0.5\cos (12\pi x/5),~\phi \left( t=0\right)
=0,  \label{input}
\end{equation}%
with parameters $G=2$, $g=1$, $q=1$, in the domain of size $L=10$. In this
case, $u_{0}^{2}=10$ is definitely smaller than critical value (\ref{cr}), $%
\left( u_{0}^{2}\right) _{\mathrm{cr}}=8\pi $. For this reason, Fig. \ref{f0}%
(a) demonstrates stable propagation of perturbations, which may be
considered as ion-acoustic waves in the cation-anion plasma, cf. Ref. \cite%
{Langmuir}. Figure \ref{f0}(b) shows the evolution of $|\psi _{+}|$ at $%
x=L/2 $. The numerically found temporal period of the wave is $T\approx 0.28$%
, which is very close to that predicted by Eq.~(\ref{g2}) at the same values
of the parameters, $2\pi /|\gamma |=0.283$.

\subsection{Density waves}

\begin{figure}[h]
\begin{center}
\includegraphics[height=5.0cm]{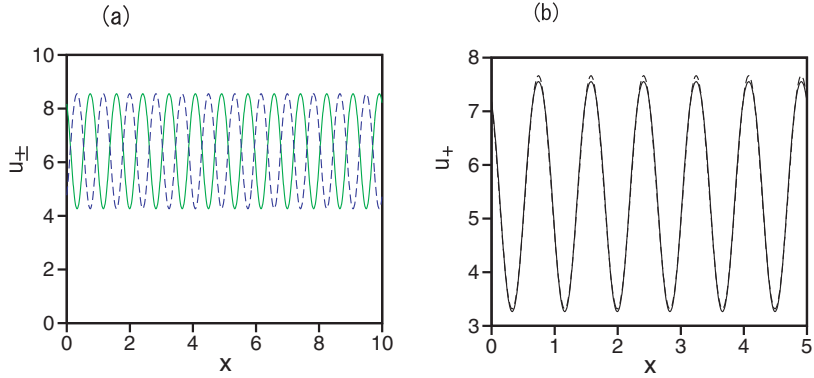}
\end{center}
\caption{(a) A stable spatially periodic numerical solution of Eqs. (\protect
\ref{+u}), (\protect\ref{-u}) and (\protect\ref{Green}) (the density wave)
with components $u_{+}(x)$ and $u_{-}(x)$ (solid green and dashed blue
lines, respectively) in the domain of size $L=10$ with periodic boundary
conditions. The chemical potential and norms of the numerical solution are $%
\protect\mu _{\pm }=94.85$ and $N_{\pm }=325$. (b) The fit of the numerical
solution for $u_{+}(x)$ (the continuous line) to the analytical ansatz, $%
u_{0}+A\cos (2\protect\pi x/l+\protect\alpha )$ (the dashed line), with $%
u_{0}=5.49$ and $A=2.17$ determined numerically as the spatial average value
of $u_{+}(x)$, and the square root of the average of $\left(
u_{+}(x)-u_{0}\right) ^{2}$. The best-fit value of the phase shift is $%
\protect\alpha =0.68$. The system's parameters are $g=1$, $G=2$, and $q=1$.}
\label{f1}
\end{figure}

Stable spatially periodic solutions were obtained as solutions of Eqs. (\ref%
{+})-(\ref{phi}), produced by means of the imaginary-time integration method
\cite{IT}, in the 1D domain of size $L$ with periodic boundary conditions.
Figure \ref{f1}(a) displays a typical example, with $L=10$ and the
density-wave's period $l=L/12$.

\begin{figure}[h]
\begin{center}
\includegraphics[height=5.0cm]{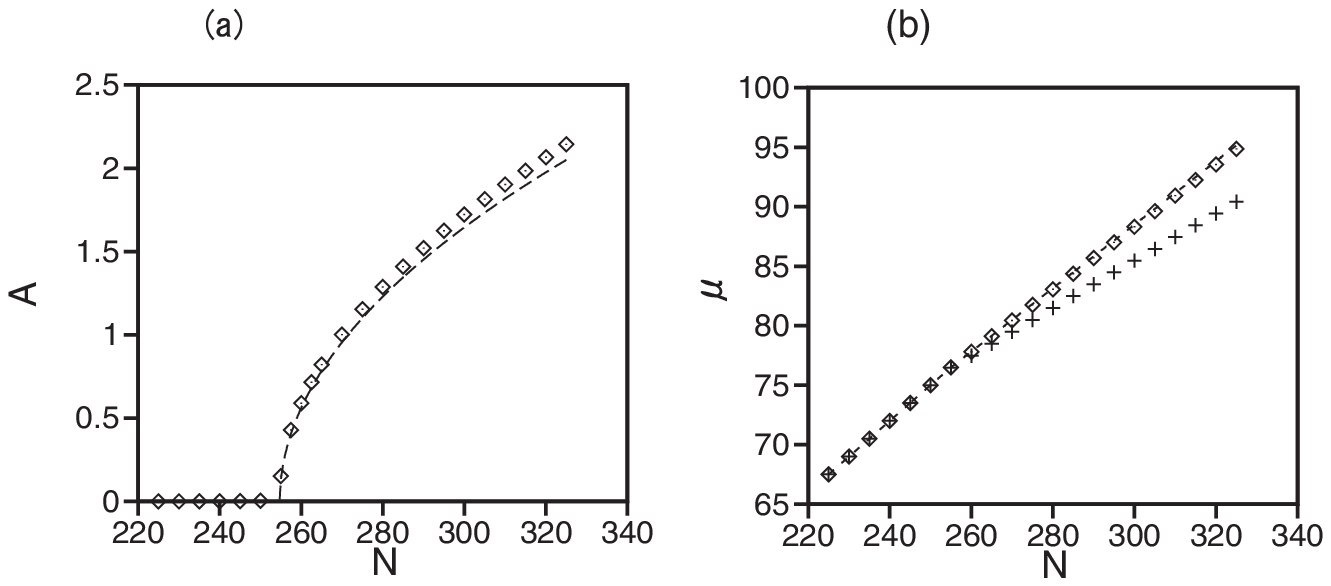}
\end{center}
\caption{(a) The modulation depth of the density wave, defined as per Eq. (
\protect\ref{Ampl}), vs. the norm of each component, $N_{\pm }=N$, for the
same system's parameter as in Fig. \protect\ref{f1}, i.e., $g=1,G=2$, $q=1$
and $L=10$. The dashed line shows $A(N)$ as predicted by analytical
approximation (\protect\ref{A}) with $l=L/12$. (b) The chain of rhombuses
and dashed line depict, severally,\ the numerically found chemical potential
of the same family of the density waves, $\protect\mu (N)$ and its
analytically predicted counterparts, given by Eq. (\protect\ref{derivative}%
). In addition, the chain of crosses shows $\protect\mu $ as predicted, in a
simple form, by Eqs. (\protect\ref{mu}) and (\protect\ref{N/L}), (\protect
\ref{A}).}
\label{f2}
\end{figure}

The pattern displayed in Fig. \ref{f1}(a) demonstrates phase separation
(demixing) of the two components, in the case when their contact repulsion,
accounted for by $G>0$ in Eqs. (\ref{+u}) and (\ref{-u}), is stronger than
the Coulomb attraction between the cations and anions. Naturally, for fixed
values of the system's parameters, \textit{viz}., $g$, $G$, $q$ in Eqs. (\ref%
{+u})-(\ref{phi}), and fixed size $L$, the destabilization of the uniform
state takes place, with the increase of the effective nonlinearity strength,
when the density of each component exceeds a certain critical value: $N_{\pm
}/L\equiv N/L>n_{\mathrm{cr}}$, see Eq. (\ref{crit}) below. Numerical
findings for the transition from the uniform neutral state to the density
wave are summarized in Fig. \ref{f2}(a), which displays the modulation depth
of the arising pattern, defined as half-difference between local maxima and
minima of $\phi _{+}(x)$,
\begin{equation}
A\equiv (1/2)\left( \phi _{\max }-\phi _{\min }\right)   \label{Ampl}
\end{equation}%
[it is the same for $\phi _{-}(x)$] as a function of $N_{\pm }\equiv N$,
with the same system's parameters as in Fig. \ref{f1}. The striped pattern
may be considered as an analog of an ionic crystal \cite{House} 
(unlike the above-mentioned Wigner crystal, which
is a pattern emerging in strongly-coupled degenerate plasmas, including the
quasi-1D setting \cite{Ichimaru,Astra-Girar}).

The critical point, $N=N_{\mathrm{cr}}\equiv n_{\mathrm{cr}}L$, at which the
periodic pattern appears, and the shape of the pattern can be predicted in
an analytical form. To this end, we note that the energy of the GPP system,
represented by Eqs. (\ref{+})-(\ref{phi}), is
\begin{equation}
E=\int_{0}^{L}\left[ \frac{1}{2}\left( \left\vert \frac{d\psi _{+}}{dx}%
\right\vert ^{2}+\left\vert \frac{d\psi _{-}}{dx}\right\vert ^{2}\right) +%
\frac{1}{2}g\left( |\psi _{+}|^{4}+|\psi _{-}|^{4}\right) +G|\psi _{+}\psi
_{-}|^{2}+\frac{1}{8\pi }\left( \frac{d\phi }{dx}\right) ^{2}\right] dx,
\label{energy}
\end{equation}%
the last term being the electrostatic energy. Then, the demixed pattern with
spatial period $l$ may be approximated by a simple variational ansatz,%
\begin{equation}
u_{\pm }(x)=u_{0}\pm A\cos \left( 2\pi x/l\right) ,  \label{u+-}
\end{equation}%
where $u_{0}$ is the mean value, and modulation depth $A$ is the same as
defined by Eq. (\ref{Ampl}), the respective density of particles being
\begin{equation}
n\equiv N/L=u_{0}^{2}+A^{2}/2.  \label{N/L}
\end{equation}%
Figure \ref{f1}(b) displays the fit of a typical numerically obtained
component $u_{+}(x)$ to ansatz $u_{0}+A\cos (2\pi x/l+\alpha )$. Further,
the substitution of ansatz (\ref{u+-}) in Eq. (\ref{phi2}) and consideration
of the balance condition for the fundamental harmonic, $\cos \left( 2\pi
x/l\right) $, yields the corresponding approximation for the electrostatic
potential:
\begin{equation}
\phi =(4/\pi )l^{2}qu_{0}A\cos \left( 2\pi x/l\right) .  \label{phi3}
\end{equation}

Next, substituting components (\ref{u+-}) and (\ref{phi3}) of the ansatz in
expression (\ref{energy}), and using Eq. (\ref{N/L}) to eliminate $u_{0}^{2}$
in favor of the density of particles, we find the corresponding energy
density,
\begin{gather}
\mathcal{E}\equiv \frac{E}{L}=(g+G)n^{2}+\left[ \frac{2\pi ^{2}}{l^{2}}+%
\frac{4}{\pi }\left( ql\right) ^{2}n+2(g-G)n\right] A^{2}  \notag \\
+\left[ \frac{9G-7g}{8}-\frac{2}{\pi }(ql)^{2}\right] A^{4}.  \label{density}
\end{gather}%
In the framework of the variational method \cite{progress}, the modulation
depth is determined by minimization of the energy density with respect to $A$%
, i.e., $d\mathcal{E}/d\left( A^{2}\right) =0$, which yields \
\begin{equation}
A^{2}=\frac{8\pi ^{2}/l^{2}+\left( 16/\pi \right) \left( ql\right)
^{2}n-8(G-g)n}{(9G-7g)-\left( 16/\pi \right) (ql)^{2}}.  \label{A}
\end{equation}

The chemical potential of the density wave can be predicted by taking Eqs. (%
\ref{+u}) and (\ref{-u}) at points where $\cos \left( 2\pi x/l\right) =0$,
hence Eqs. (\ref{u+-}) and (\ref{phi3}) give $u_{+}=u_{-}=u_{0}$ and $%
d^{2}u_{\pm }/dx^{2}=\phi =0$:
\begin{equation}
\mu _{\pm }=\left( G+g\right) u_{0}^{2}.  \label{mu}
\end{equation}%
Figure \ref{f2}(b) shows the numerically found chemical potential $\mu $
(rhombuses) with $\mu $ predicted by Eq.~(\ref{mu}) (crosses) and,
additionally, the chemical potential calculated, as usual, as the derivative
of the energy density with respect to the total density ($2n$):
\begin{equation}
\mu =d(\mathcal{E})/d(2n),  \label{derivative}
\end{equation}
obtained from Eq.~(\ref{density}) and depicted by the dashed line.

The result given by Eq. (\ref{A}) makes sense at $A^{2}\geq 0$, i.e., at
\begin{equation}
n\geq n_{\mathrm{cr}}=8\pi q^{2}/\left( G-g\right) .  \label{crit}
\end{equation}%
At the critical point, $n=n_{\mathrm{cr}}$, the MI is driven by
perturbations with spatial period%
\begin{equation}
l_{\mathrm{cr}}=\left( \sqrt{\pi }/2\right) q^{-1}.  \label{lcr}
\end{equation}%
In fact, Eqs. (\ref{crit}) and (\ref{lcr}) yield exact results, which are
identical, respectively, to Eqs. (\ref{cr}) and (\ref{pcr}), with $l_{%
\mathrm{cr}}\equiv 2\pi /p_{\mathrm{cr}}$. This is explained by the fact
that the ansatz based on Eqs. (\ref{u+-}) and (\ref{phi3}) with
infinitesimal $A$ reproduces the exact MI\ eigenmode which leads to Eqs. (%
\ref{g1})-(\ref{max}). In particular, at $q=1$ and $L=10$, Eq. (\ref{crit})
yields $N_{\mathrm{cr}}\equiv Ln_{\mathrm{cr}}\approx 251.3$, which
precisely agrees with the numerically found critical value of $N$ in Fig. %
\ref{f2}(a).

For the same case, $q=1$, Eq. (\ref{lcr}) yields $l_{\mathrm{cr}}=\sqrt{\pi }%
/2\approx L/\allowbreak 11.284$ for $L=10$ [the calculation of the critical
value (\ref{lcr}) is not subject to the condition that ratio $L/l_{\mathrm{cr%
}}$ must be integer]. Picking up a close integer, $12$, i.e., $l=L/12$, the
dashed line in Fig. \ref{f2}(a) shows $A$, as predicted by Eq.~(\ref{A}),
vs. $N $ for $l=L/12$ and fixed $q=1$, $G=2$, $g=1$. The figure demonstrates
very accurate agreement of the analytically predicted $A(N)$ dependence with
the numerically found counterpart.

It is relevant to compare energies $E$ of different stationary states
sharing the same value of $N$. The calculation of $E$ as per Eq. (\ref%
{energy}) demonstrates that, at $N>N_{\mathrm{cr}}$, the energy is always
lower for the density-wave pattern than for the uniform (neutral) state
existing at the same value of $N$. Therefore, it is plausible that the
spatially periodic pattern realizes the system's GS.

In addition to the above family of spatially-periodic solutions with two
free parameters, $l$ and $n$, found in the numerical and approximate
analytical forms, it is possible to find a family of \emph{exact analytical
solutions}, in the form given by Eq. (\ref{u}), with

\begin{gather}
u_{+}(x)=U_{0}\cos (2\pi x/l_{0}),~u_{-}=U_{0}\sin (2\pi x/l_{0}),~  \notag
\\
\phi =\Phi _{0}\cos (4\pi x/l_{0}),  \notag \\
\Phi _{0}=(G-g)U_{0}^{2}/(2q),~l_{0}=\sqrt{2\pi (G-g)}q^{-1},  \notag \\
\mu _{\pm }=\frac{2\pi ^{2}}{l_{0}^{2}}+\frac{1}{2}\left( G+g\right)
U_{0}^{2}~.  \label{exact}
\end{gather}
The norms of solution (\ref{exact}) are
\begin{equation}
N\equiv N_{\pm }=U_{0}^{2}L/2.  \label{Nexact}
\end{equation}%
Amplitude $U_{0}$ is a single free parameter of the family, while the
spatial period takes the single value, $l_{0}$.

\begin{figure}[h]
\begin{center}
\includegraphics[height=5.0cm]{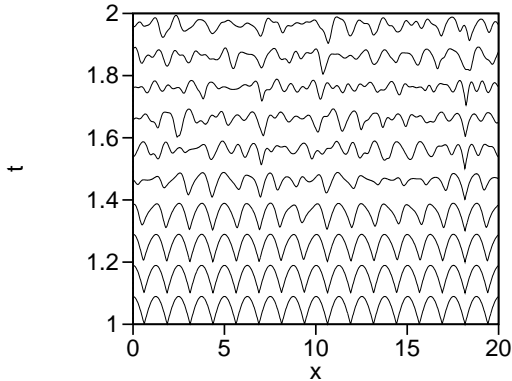}
\end{center}
\caption{The perturbed evolution initiated by exact solution (\protect\ref%
{exact}) with $U_{0}=2\protect\sqrt{5}$. Snapshots of $\left\vert \protect%
\psi _{+}(x,t)\right\vert $ are plotted at $t=1+0.1n$ with $n=0,1,\cdots 9$.
The parameters are $g=1,G=2$, and $q=1$. The simulations were performed in
the domain of size $L=8\protect\sqrt{2\protect\pi }\equiv 8l_{0}$ [see Eq. (%
\protect\ref{exact})] with periodic boundary conditions.}
\label{f3}
\end{figure}

On the contrary to the above numerical solutions, the exact ones (\ref{exact}%
) are unstable, as illustrated by simulations of the evolution displayed in
Fig. \ref{f3} for the exact solution with amplitude $U_{0}=2\sqrt{5}$. The
simulations were run in the domain of size $L=8l_{0}$, to make $l_{0}$
compatible with the periodic boundary conditions. The instability, which
leads to establishment of an apparently turbulent state, is explained by the
fact that the energy of exact solution (\ref{exact}) is larger than that of
the uniform neutral state with the same norm:%
\begin{equation}
E_{\mathrm{exact}}=(G+g)N^{2}/L+2\pi N(G-g)>E_{\mathrm{neutral}}=\left(
G+g\right) N^{2}/L,
\end{equation}%
i.e., the exact solution represents an excited state of the system. Note
that, for these values of the parameters, norm (\ref{Nexact}) is $N=80\sqrt{%
2\pi }\approx \allowbreak 200.53$, being smaller than the respective
critical value, $N_{\mathrm{cr}}=8\pi q^{2}L\approx 503.\allowbreak 99$, as
given by Eqs. (\ref{cr}) and (\ref{crit}), hence the uniform neutral state
with the same $N$ is indeed stable, representing the system's GS.

\begin{figure}[h]
\begin{center}
\includegraphics[height=5.0cm]{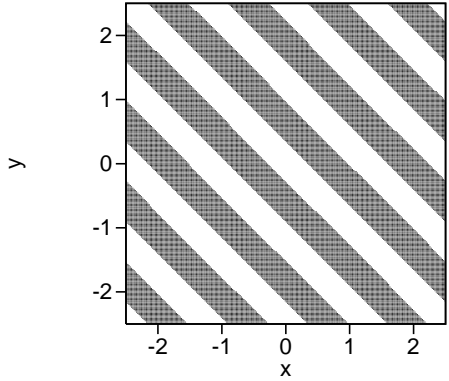}
\end{center}
\caption{A stable 2D striped pattern with norms $N_{+}=N_{-}=650$, produced
by the imaginary-time solution of Eqs. (\protect\ref{+})-(\protect\ref{phi})
in a square of size $5\times 5$ with periodic boundary conditions, see the
definition of black stripes given by Eq. (\protect\ref{>}). The parameters
are $G=2$ and $g=q=1$.}
\label{f4}
\end{figure}

Numerical solution of the 2D version of Eqs. (\ref{+})-(\ref{phi}) readily
produces stable spatially periodic patterns in the form of quasi-1D stripes,
see an example in Fig. \ref{f4}, in which black areas are defined as those
with%
\begin{equation}
\left\vert \psi \left( x,y\right) \right\vert \geq (1/2)\left[ \left(
\left\vert \psi (x,y\right\vert \right) _{\max }+\left( \left\vert \psi
(x,y\right\vert \right) _{\min }\right] ,  \label{>}
\end{equation}%
where subscripts $\max $ and $\min $ denote, severally, the largest and
smallest values of $\left\vert \psi (x,y)\right\vert $ as functions of $x$
and $y$. Integral norms (\ref{N}) of this state, in the 2D area of size $%
5\times 5$,\ are $N_{+}=N_{-}=650$, hence the corresponding average
densities are $n=N_{\pm }/25=\allowbreak 26$. Note that the above MI
analysis pertains equally well to all spatial dimensions. For the present
values of the parameters ($G=2$, $g=q=1$), the critical density, given by
Eqs. (\ref{cr}) and (\ref{crit}), is $n_{\mathrm{cr}}=8\pi \approx
\allowbreak 25.133$. The existence of the stable striped pattern is natural,
as $n=26$ exceeds $n_{\mathrm{cr}}$, hence the uniform neutral solution
should be replaced by the striped one, as the respective GS.

Finally, Eqs. (\ref{+})-(\ref{phi}) admit an exact 2D\ solution in the form
of a square-lattice pattern,
\begin{gather}
\psi _{+}\left( x,y\right) =e^{-i\mu t}\left[ U_{0x}\cos (2\pi
x/l_{0})+ie^{-i\mu t}U_{0y}\cos (2\pi y/l_{0})\right] ,  \label{psi+2D} \\
\psi _{-}\left( x,y\right) =e^{-i\mu t}\left[ U_{0x}\sin (2\pi
x/l_{0})+ie^{-i\mu _{y}t}U_{0y}\sin (2\pi y/l_{0}\right] ),  \label{psi-2D}
\\
\phi (x)=\Phi _{0x}\cos \left( 4\pi x/l_{0}\right) +\Phi _{0y}\cos (4\pi
y/l_{0}),  \label{phi2D}
\end{gather}%
with the same single value $l_{0}$ as given above by Eq. (\ref{exact}), real
amplitudes $U_{0x,0y}$ and $\Phi _{0x,0y}$, which are coupled by relations
\begin{equation}
\Phi _{0x,0y}=\frac{G-g}{2q}\left( U_{0x,0y}\right) ^{2}~,  \label{phipsi2D}
\end{equation}%
and chemical potential
\begin{equation}
\mu =\frac{2\pi ^{2}}{l_{0}^{2}}+\frac{1}{2}(G+g)\left(
U_{0x}^{2}+U_{0y}^{2}\right) .  \label{mu2D}
\end{equation}%
In the 2D area of size $L\times L$, the norms of the exact 2D solution is $%
N_{\pm }=\left( U_{0x}^{2}+U_{0y}^{2}\right) L^{2}/2$. Similar to the exact
1D solutions, the 2D ones are unstable in direct simulations (not shown here
in detail).

\section{Solitons}

\subsection{Dipole modes: analytical and numerical results.}

The underlying system of Eqs. (\ref{+})-(\ref{phi}) may give rise to bright
solitons in the case of the attractive sign of the intra-component contact
nonlinearity, i.e., $g<0$. It is expected that the Coulomb attraction will
tend to keep the two components together, in competition with the contact
repulsion between them, accounted for by $G>0$. Thus, creation of a
dipole-shaped soliton is expected.

\begin{figure}[h]
\begin{center}
\includegraphics[height=5.0cm]{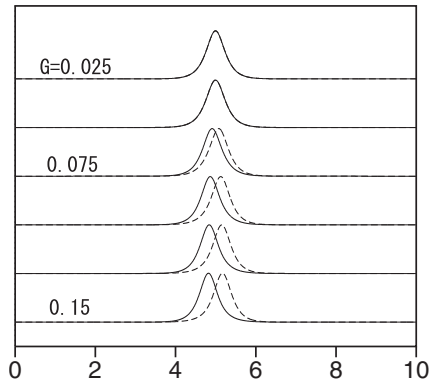}
\end{center}
\caption{Stationary profiles of $u_{+}(x)$ and $u_{-}(x)$ (solid and dashed
lines, respectively) of stable dipole-shaped solitons, produced as solutions
of Eqs. (\protect\ref{+})-(\protect\ref{phi}) in the domain of size $L=10$
with Neumann's boundary conditions. The profiles are shown at a set of
decreasing values of the contact-repulsion strength, $%
G=0.15,0.125,0.1,0.075,0.05$, and $0.025$. All solitons have norms $N_{\pm
}=10$. Other parameters in Eqs. ( \protect\ref{+})-(\protect\ref{phi}) are $%
g=-1$ and $q=0.3$.}
\label{f5}
\end{figure}

Figure \ref{f5} shows a set of stable solitons, obtained by means of the
imaginary-time-integration method in the domain of size $L=10$ with
zero-flux (Neumann's) boundary conditions and fixed norms, $N_{\pm }=10$. It
is seen that $L=10$ is sufficient to produce completely localized states.
For fixed parameters $g=-1$, $q=0.3$ and varying $G$, the two components
remain fully overlapped at $G\leq 0.05$, splitting at $G>0.05$. In the state
with coinciding components, an obvious exact soliton solution with the
center set at $x=L/2$ (see Fig. \ref{f5}) is
\begin{equation}
\psi _{\pm }=\frac{1}{\sqrt{|g|-G}}\frac{A}{\cosh \left( Ax^{\prime }\right)
}\exp \left( -\frac{i}{2}A^{2}t\right) ,~\phi =0.  \label{AA}
\end{equation}%
provided that $G<$ $|g|$. Here, $x^{\prime }\equiv x-L/2$, and $A$ is an
arbitrary amplitude of the soliton.

To approximate solutions with small separation $2\xi $ between the
components, we adopt a variational ansatz suggested by solution \ref{AA}:
\begin{equation}
\psi _{\pm }=\frac{1}{\sqrt{|g|-G}}\frac{A}{\cosh \left[ A(x^{\prime }\mp
\xi )\right] }\exp \left( -\frac{i}{2}A^{2}t\right) ,  \label{ansatz}
\end{equation}%
with norms
\begin{equation}
N_{\pm }=2A/\left( |g|-G\right) .  \label{NN}
\end{equation}%
The dipole moment of the weakly split state represented by ansatz (\ref%
{ansatz}) is%
\begin{equation}
\mathrm{DM}\equiv \int_{-L/2}^{+L/2}x^{\prime }\left[ \left\vert \psi
_{+}(x^{\prime })\right\vert ^{2}-\left\vert \psi _{-}(x^{\prime
})\right\vert ^{2}\right] dx^{\prime }\approx \frac{4A\xi }{|g|-G}.
\label{dip}
\end{equation}%
The respective Poisson equation (\ref{phi}) takes the form of
\begin{equation}
\frac{d^{2}\phi }{dx^{2}}=-\frac{4\pi qA^{2}}{|g|-G}\left\{ \frac{1}{\cosh
^{2}\left[ A(x^{\prime }-\xi )\right] }-\frac{1}{\cosh ^{2}\left[
A(x^{\prime }+\xi )\right] }\right\} .  \label{phi4}
\end{equation}%
Straightforward integration of Eq. (\ref{phi4}) yields
\begin{equation}
\frac{d\phi }{dx}=-\frac{4\pi qA}{|g|-G}\left\{ \tanh \left[ A(x^{\prime
}-\xi )\right] -\tanh \left[ A(x^{\prime }+\xi )\right] \right\} .
\label{tanh}
\end{equation}%
The Coulomb energy of the soliton with the slightly split components can be
readily evaluated, using approximation (\ref{tanh}):%
\begin{gather}
E_{\phi }=\frac{1}{8\pi }\int_{0}^{L}\left( \frac{d\phi }{dx}\right)
^{2}dx\approx \frac{2\pi q^{2}A^{2}}{(|g|-G)^{2}}\int_{-\infty }^{+\infty
}\left\{ \tanh \left[ A(x^{\prime }-\xi )\right] -\tanh \left[ A(x^{\prime
}+\xi )\right] \right\} ^{2}dx^{\prime }  \notag \\
\approx \frac{32\pi q^{2}A^{3}}{3(|g|-G)^{2}}\xi ^{2}\left( 1-\frac{4}{15}%
A^{2}\xi ^{2}\right) ,  \label{E2}
\end{gather}%
where the expansion is built for small $\xi ^{2}$.

Further, the energy of the contact interaction of the two components in
ansatz (\ref{ansatz}) is found as
\begin{gather}
E_{G}=G\int_{0}^{L}|\psi _{+}(x)|^{2}|\psi _{-}(x)|^{2}dx\approx \frac{GA^{4}%
}{(|g|-G)^{2}}\int_{-\infty }^{+\infty }\frac{1}{\cosh ^{2}\left(
A(x^{\prime }-\xi )\right) }\frac{1}{\cosh ^{2}\left( A(x^{\prime }+\xi
)\right) }dx^{\prime }  \notag \\
\approx \frac{4GA^{4}}{3(|g|-G)^{2}}\left( 1-\frac{8}{5}A^{2}\xi ^{2}+\frac{%
32}{21}A^{4}\xi ^{4}\right) .  \label{E1}
\end{gather}%
Then, the total energy of the interaction of the two weakly separated
components is
\begin{equation}
E_{\phi }+E_{G}=\frac{4A^{3}}{3(|g|-G)^{2}}\left[ G+8\left( \pi q^{2}-\frac{1%
}{5}GA^{2}\right) \xi ^{2}+\left( \frac{32}{21}GA^{2}-\frac{32}{15}\pi
q^{2}\right) A^{2}\xi ^{4}\right] .  \label{int}
\end{equation}

\begin{figure}[h]
\begin{center}
\includegraphics[height=5.0cm]{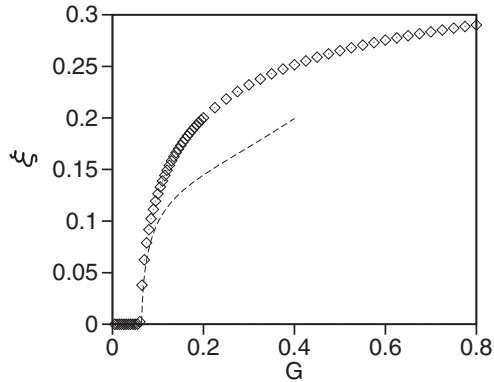}
\end{center}
\caption{Numerically found values (rhombuses) of the half-distance between
centers of the $|\protect\psi _{+}|^{2}$ and $|\protect\psi _{-}|^{2}$
components in the soliton. The dashed line shows the analytical prediction
for the same separation, produced by Eq. (\protect\ref{xiGS}), with $A^{2}$
expressed in terms of $N_{\pm }$ by means of Eq. (\protect\ref{NN}).
Parameters are $g=-1$ and $q=0.3$. The norms are $N_{\pm }=10$, and the size
of the domain of the numerical solution is $L=10$.}
\label{f6}
\end{figure}

Thus, the value of separation $\xi $ is predicted as one for which the total
interaction energy attains a minimum, $\partial \left( E_{\phi
}+E_{G}\right) /\partial \left( \xi ^{2}\right) =0$, the result being
\begin{equation}
\xi ^{2}=\frac{21\left( GA^{2}-5\pi q^{2}\right) }{8A^{2}\left( 5GA^{2}-7\pi
q^{2}\right) }  \label{xiGS}
\end{equation}%
at $A^{2}>5\pi q^{2}/G$, and $\xi ^{2}=0$ at $A^{2}>5\pi q^{2}/G$. The
dependence of $\xi $ on $G$, as produced by Eq. (\ref{xiGS}), along with its
numerically found counterpart, is shown in Fig. \ref{f6}. The weak splitting
of the components, with small $\xi $, is well predicted by the analytical
approximation, while the simple approximation used above becomes irrelevant
at larger values of the separation. In particular, for given $G<|g|$, the
splitting takes place provided that
\begin{equation}
q^{2}<q_{\mathrm{cr}}^{2}=\frac{N^{2}G}{20\pi }\left( |g|-G\right) ^{2},
\label{qcr}
\end{equation}%
where Eq. (\ref{NN}) is used to eliminate $A^{2}$ in favor of the physically
relevant norm parameter. At $q^{2}=q_{\mathrm{cr}}^{2}$, Eq. (\ref{xiGS})
gives $\xi =0$. Equation (\ref{qcr}) may be easily inverted, to produce a
critical value of $G$ at which the splitting sets in, for given $q$.

\subsection{Quadrupole solitons}

The system under the consideration gives rise not only to dipole solitons,
but also to quadrupoles, in which one component is located at the center,
while the other one splits in two side lobes. The respective quadrupole
moment is%
\begin{equation}
\mathrm{QM}=2\int_{-L/2}^{+L/2}(x^{\prime })^{2}\left[ \left\vert \psi
_{+}(x^{\prime })\right\vert ^{2}-\left\vert \psi _{-}(x^{\prime
})\right\vert ^{2}\right] dx^{\prime },  \label{quad}
\end{equation}%
cf. Eq. (\ref{dip}). The simplest ansatz which may approximate quadrupole
solitons is%
\begin{equation}
\psi _{\pm }=\frac{1}{\sqrt{|g|-G}}\left[ 1\pm \varepsilon (Ax^{\prime
})^{2}\mp \frac{\pi ^{2}}{12}\varepsilon \right] \frac{A}{\cosh \left(
Ax^{\prime }\right) }\exp \left( -\frac{i}{2}A^{2}t\right) ,  \label{Qans}
\end{equation}%
with small $\varepsilon >0$. The norm of ansatz (\ref{Qans}) keeps value (%
\ref{NN}) at order $\varepsilon $. The quadrupole moment produced by the
substitution of ansatz (\ref{Qans}) in Eq. (\ref{quad}) is $\mathrm{QM}%
=\left( 7\pi ^{4}/15\right) A^{-1}\left( |g|-G\right) ^{-1}\varepsilon +%
\mathcal{O}\left( \varepsilon ^{2}\right) $.

\begin{figure}[h]
\begin{center}
\includegraphics[height=4.0cm]{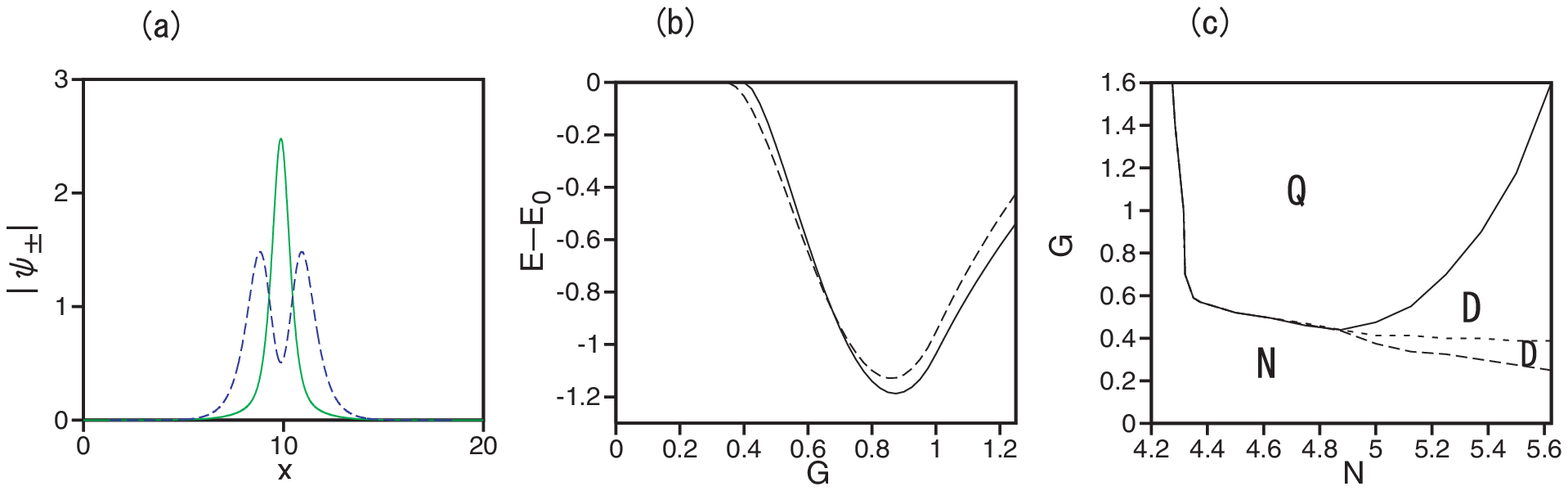}
\end{center}
\caption{ (a) Stationary profiles of $\left\vert \protect\psi %
_{+}\right\vert $ and $\left\vert \protect\psi _{-}\right\vert _{-}$ (solid
green and dashed blue lines, respectively) for a stable quadruple soliton
with norms $N_{\pm }=5$, obtained by means of the imaginary-time method as a
solution of Eqs. (\protect\ref{+})-(\protect\ref{phi}) in the domain of size
$L=20$, with the Neumann's boundary conditions. The parameters are $g=-1$, $%
G=0.9$, and $q=0.3$. The quadrupole moment of the soliton [see Eq. (\protect
\ref{quad})] is $\mathrm{QM}=-13.54$. (b) The energy difference, $E-E_{0}$,
between quadrupole and dipole solitons (solid \ and dashed lines,
respectively) and the neutral one (\protect\ref{AA}), for fixed norms, $%
N_{\pm }=5.25$, vs. $G$, keeping $g=-1$ and $q=0.3$ fixed too. The GS
(ground state) is identified as one with the lowest energy. (c) The chart
which identifies GS as different soliton species (Q, D, and N: quadrupole,
dipole, and neutral ones, respectively) in the plane of $N\equiv N_{\pm }$
and $G$, for fixed $g=-1$ and $q=0.3$. In region D above the short-dashed
line in Fig. \protect\ref{f7}(c), unstable quadrupole solitons also exist,
in addition to the stable dipoles.}
\label{f7}
\end{figure}

An example of a numerically obtained stable quadrupole soliton is shown in
Fig. \ref{f7}(a) for $G=0.9$, $g=-1$, $q=0.3$. Note that ansatz (\ref{Qans})
gives rise to a minimum of the $\psi _{-}$ component at $x^{\prime }=0$
and a pair of adjacent maxima at $\varepsilon >1/2$ (the quadrupole's
shape is somewhat similar to that of dark solitons predicted for a quantum
electron plasma in Ref. \cite{Shukla}). With these parameters,
quadrupoles are stable in the interval of values of their norms $4.4\leq
N_{\pm }\leq 5.6$. To identify the system's GS, Fig. \ref{f7}(b) shows the
energy difference between the dipole and quadrupole solitons and the neutral
one, given by Eq. (\ref{AA}). The solitons with equal norms, $N_{\pm }=5.25$%
, are compared here. In Fig. \ref{f7}(a), the energy differences are shown
as functions of the strength of the inter-component repulsion, $G$. It is
seen that neutral, dipole, and quadrupole solitons realize the GS (energy
minimum) at $G<0.4$, $0.4<G<0.7$, and $G>0.7$, respectively.

The results are summarized in the chart displayed in Fig. \ref{f7}(c), which
identifies the system's GS as neutral (N), dipole (D), or quadrupole (Q)
solitons, in the parameter plane of $N_{\pm }$ and $G$, with fixed $g=-1$
and $q=0.3$. In the figure, the neutral solitons lose their stability and
give rise, by splitting, to the quadrupole and dipole solitons at boundaries
between the N and Q or D areas, respectively (at the boundaries, the
solitons of different types have equal energies). It is natural that the
increase of the repulsion strength, $G$, drives splitting of neutral
solitons into quadrupole and dipole ones in Fig. \ref{f7}(c). It is natural
too that further increase of $G$ leads to additional fragmentation,
replacing the dipole solitons by quadrupole ones as the GS.

\begin{figure}[h]
\begin{center}
\includegraphics[height=5.5cm]{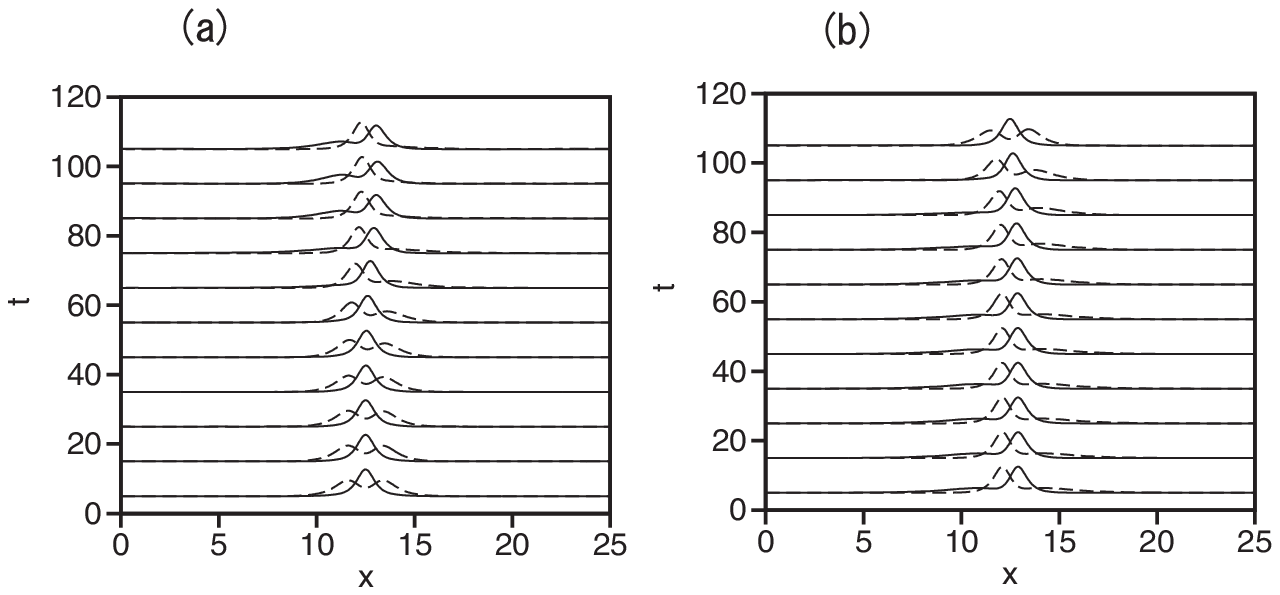}
\end{center}
\caption{ (a) Spontaneous transformation of an unstable quadrupole soliton
with norm $N_{\pm }=5.25$ into a stable dipole one, at $g=-1,$ $q=0.3$, and $%
G=0.65$. (b) The transformation of an unstable dipole into a stable
quadrupole at the same values of $\ q$, $g$, and $N_{\pm }$, but with larger
$G=0.75$. In terms of the chart in Fig. \protect\ref{f8}(c), parameters
corresponding to cases (a) and (b) belong to areas D (above the internal
short-dashed boundary) and Q (slightly above its bottom boundary),
respectively. The evolution is displayed by means of profiles of $\left\vert
\protect\psi _{\pm }\left( x,t\right) \right\vert $.}
\label{f8}
\end{figure}

No region of coexistence of stable solitons of different types could be
found. In particular, in the part of area D above the short-dashed line in
Fig. \ref{f7}(c), quadrupole solitons exist but are unstable, spontaneously
transforming into dipole counterparts, see an illustration in Fig. \ref{f8}%
(a). In the Q area, dipole solitons exist too, but they are unstable against
spontaneous transformation into quadrupole ones, see Fig. \ref{f8}(b).

\subsection{Collisions between traveling dipole solitons}

\begin{figure}[h]
\begin{center}
\includegraphics[height=5.5cm]{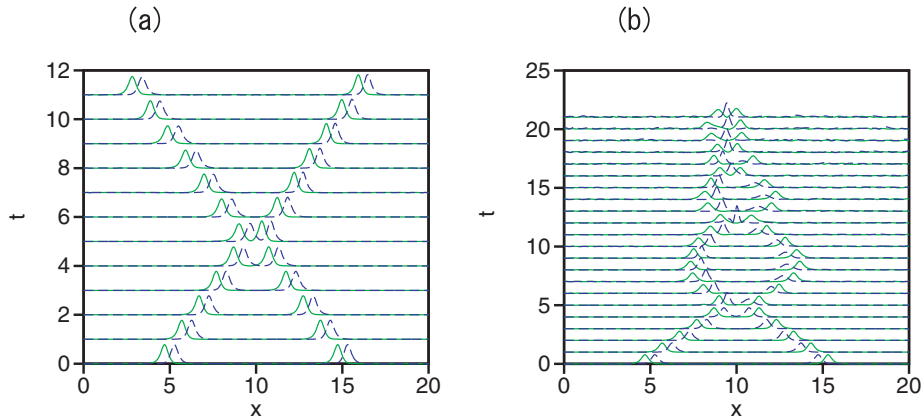}
\end{center}
\caption{Panels (a) and (b) display, severally, typical examples of
dipole-dipole and dipole-antidipole soliton-soliton collisions, with
velocities $c=\pm 1$. In both cases, norms of the solitons are $N_{\pm }=10$
and the size of the domain is $L=20$. The parameters in Eqs. (\protect\ref{+}%
)-(\protect\ref{phi}) are $G=0.9$, $g=-1$, and $q=0.3$. The dipole-dipole
collision is elastic, while the dipole-antidipole interaction leads to
recurring collisions and, eventually, merger of the dipole-antidipole pair
into a stable quadrupole soliton.}
\label{f9}
\end{figure}

The Galilean invariance of the underlying system of Eqs. (\ref{+})-(\ref{phi}%
) makes it possible to set stable solitons in motion with velocity $c$, by
applying a kick to them, $\psi _{\pm }\rightarrow \psi _{\pm }e^{icx}$. This
way, collisions between solitons moving with velocities $\pm c$ may be
simulated. We have considered two types of collisions of dipole solitons,
with identical or opposite dipole moments, i.e., $\left( \mathrm{DM},\mathrm{%
DM}\right) $ or $\left( \mathrm{DM},-\mathrm{DM}\right) $. Obviously, the
dipole-dipole interaction force is attractive in the former case, and
repulsive in the latter one. Figure \ref{f9}(a) displays a typical example
of the collision in the former case, with velocities $c=\pm 1$, in the
domain of size $L=20$, for $g=-1$, $G=0.9$, $L=20$, and $N_{\pm }=10$. The
collision is elastic, with the solitons readily passing through each other.
On the other hand, Fig. \ref{f9}(b) demonstrates that, for the same
parameters, the solitons with opposite dipole moments undergo multiple
collisions (ca. four in this picture), bouncing back and colliding again,
until merger of the dipole-antidipole pair into a stable quadrupole soliton.
It features a central $\psi _{-}$ component sandwiched between two $\psi _{+}
$ side lobes, cf. Fig. \ref{f7}(a).

\section{Conclusion}

The objective of this work is to introduce 1D and 2D models of the
\textquotedblleft ultracold plasma", composed of two atomic species, which
carry positive and negative charges (cations and anions), interacting via
the electrostatic field governed by the Poisson equation. An estimate of
parameters for a relatively dense degenerate plasma composed of light
cations and anions demonstrates that, in the \textquotedblleft bare form",
it would be too strongly coupled by the Coulomb interaction; however,
reduction of the effective mass by means of a lattice potential may bring
the system into the mean-field state described by the GP equations for wave
functions of the ionic species, coupled to the Poisson equation for the
electrostatic field. The GP equations include, along with the Coulomb terms,
contact interactions, namely, the repulsion between the species and
self-repulsion or attraction in each one. The contact interactions may be
included directly, or induced by a buffer species of heavy neutral atoms
(which may also be used for sympathetic cooling of the ions). The estimate
demonstrates that, in the above-mentioned regime with the reduced effective
mass, the contact interactions provide for relevant competition with the
Coulomb forces and kinetic energy of the particles. For the system with full
contact repulsion, the main objective is to predict stable spatially
periodic density-wave patterns, which are akin to ionic crystals
in solid-state physics. The
transition from the uniform neutral state to the density wave is exactly
identified by means of the analytical consideration of the MI (modulational
instability) of the uniform state. The emerging 1D pattern is accurately
predicted by means of the variational approximation, and found in the
numerical form. It represents the system's GS above the onset of the MI. In
the 2D setting, a stable density wave is found with the quasi-1D shape.

The 1D system with contact self-attraction in each ionic species gives rise
to bright solitons. If the contact repulsion between the species is strong
enough, neutral solitons split into dipole or quadrupole states, which
represent the GS. The transition from the neutral solitons to dipole ones is
accurately predicted by an analytical approximation. Different types of the
solitons may coexist as stationary states, but only one of them is stable.
Collisions between moving dipole solitons were simulated too. The result is
that the dipole-dipole collision is elastic, while the dipole-antidipole
pair features multiple collisions, eventually merging into a stable
quadrupole soliton.

This work can be developed in other directions. In particular, as mentioned
above, it may be relevant to add a wave function representing polar
molecules built as a cation-anion bound states. The respective GP equations
should then include \textquotedblleft reactions" between colliding ions and
molecules. A challenging extension is to address the 3D version of the
present system, and to study the 2D case more systematically. In particular,
it may be relevant to construct domain-wall patterns between two sets of
stable stripes, such as those displayed in Fig. \ref{f4}, but with different
orientations \cite{old,Staliunas}.

\section*{Acknowledgments}

The work of B.A.M. is supported, in part, by the Israel Science Foundation
through grant No. 1287/17. This author appreciates hospitality of the
Interdisciplinary Graduate School of Engineering Sciences at the Kyushu
University (Fukuoka, Japan). The work of H.S. is supported by the Japan
Society for Promotion of Science through KAKENHI Grant No. 18K03462.

\section*{Appendix}

The dispersion equation for modulational perturbations, taken as per Eq. (%
\ref{pert}) and substituted in the linearized version of Eqs. (\ref{+})-(\ref%
{phi}), is derived in the following determinant form:

\begin{equation}
\left\vert
\begin{array}{ccccc}
\gamma & -\frac{1}{2}p^{2} & 0 & 0 & 0 \\
-\frac{1}{2}p^{2}-2gu_{0}^{2} & -\gamma & -2Gu_{0}^{2} & 0 & -qu_{0} \\
0 & 0 & \gamma & -\frac{1}{2}p^{2} & 0 \\
-2Gu_{0}^{2} & 0 & -\frac{1}{2}p^{2}-2gu_{0}^{2} & -\gamma & qu_{0} \\
8\pi qu_{0} & 0 & -8\pi qu_{0} & 0 & -p^{2}%
\end{array}%
\right\vert =0.
\end{equation}%
The calculation of the determinant leads to the expressions for $\gamma
(p)$ displayed in Eqs. (\ref{g1}) and (\ref{g2}).


\begin{thebibliography}{99}
\bibitem{Anderson} M. H. Anderson, J. R. Ensher, M. R. Matthews, C. E.
Wieman, and E. A. Cornell, Observation of Bose-Einstein condensation in a
dilute atomic vapor, Science \textbf{269}, 198-201 (1995).

\bibitem{GP} L. Pitaevskii and S. Stringari, \textit{Bose-Einstein
Condensation }(Clarendon Press, Oxford, 2003).

\bibitem{dipole-review1} M. A. Baranov, Theoretical progress in many-body
physics with ultracold dipolar gases, Phys. Rep. \textbf{464}, 71-111 (2008).

\bibitem{dipole-review2} T. Lahaye, C. Menotti, L. Santos, and M.
Lewenstein, and T. Pfau, The physics of dipolar bosonic quantum gases, Rep.
Prog. Phys. \textbf{72}, 126401 (2009).

\bibitem{Bao} W. Bao, Y. Yongyong, and H. Wang, Efficient numerical methods
for computing ground states and dynamics of dipolar Bose-Einstein
condensates, J. Comput. Phys. \textbf{229}, 7874-7892 (2010).

\bibitem{dipole-review3} C. Trefzger, C. Menotti, B. Capogrosso-Sansone, and
M. Lewenstein, Ultracold dipolar gases in optical lattices, J. Phys. B: At.
Mol. Opt. Phys. \textbf{44}, 193001 (2011).

\bibitem{Yukalovs} V.~I.~Yukalov and E.~P.~Yukalova, Bose-condensed atomic
systems with nonlocal interaction potentials, Laser Phys. \textbf{26},
045501 (2016).

\bibitem{gravity1} P.-H. Chavanis, Mass-radius relation of Newtonian
self-gravitating Bose-Einstein condensates with short-range interactions. I.
Analytical results, Phys. Rev. D \textbf{84}, 043531 (2011).

\bibitem{gravity2} P.-H. Chavanis and L. Delfini, Mass-radius relation of
Newtonian self-gravitating Bose-Einstein condensates with short-range
interactions. II. Numerical results, Phys. D \textbf{84}, 043532 (2011).

\bibitem{monthly1} T. Harko, Evolution of cosmological perturbations in
Bose-Einstein condensate dark matter, Monthly Not. Roy. Astron. Soc. \textbf{%
413}, 3095-3104 (2011).

\bibitem{monthly2} T. Rindler-Daller and P. R. Shapiro, Angular momentum and
vortex formation in Bose-Einstein-condensed cold dark matter haloes, Monthly
Not. Roy. Astron. Soc. \textbf{422}, 135-161 (2012).

\bibitem{gravity3} E. Madarassy, J. M. Enik, and V. T. Toth, Evolution and
dynamical properties of Bose-Einstein condensate dark matter stars, Phys.
Rev. D \textbf{91}, 044041 (2015).

\bibitem{gravity4} F. S. Guzman, Oscillation modes of ultralight BEC dark
matter cores, Phys. Rev. D \textbf{99}, 083513 (2019).

\bibitem{Kurizki} D. O'Dell, S. Giovanazzi, G. Kurizki, and V. M. Akulin,
Bose-Einstein condensates with $1/r$ interatomic attraction:
Electromagnetically induced ``gravity", Phys. Rev. Lett.\textbf{\ 84},
5687-5690 (2000).

\bibitem{gravity emulation} C. Barcelo, S. Liberati, and M. Visser, Analogue
gravity from Bose-Einstein condensates, Class. Quant. Gravity \textbf{18},
1137-1156 (2001).

\bibitem{we} H. Sakaguchi and B. A. Malomed, Suppression of the
quantum-mechanical collapse by repulsive interactions in a quantum gas,
Phys. Rev. A \textbf{83}, 013607 (2011).

\bibitem{Killian-PRL} T. C. Killian, S. Kulin, S. D. Bergeson, L. A. Orozco,
C. Orzel, and S. L. Rolston, Creation of an ultracold neutral plasma, Phys.
Rev. Lett. \textbf{83}, 4776-4779 (1999).

\bibitem{Kira} M. Kira and S. W. Koch, Many-body correlations and excitonic
effects in semiconductor spectroscopy, Prog. Quant. Electr. \textbf{30},
155-296 (2006).

\bibitem{Killian} T. C. Killian, T. Pattard, T. Pohl, and J. M. Rost,
Ultracold neutral plasmas, Phys. Rep. \textbf{449}, 77-130 (2007).

\bibitem{Rolston-earlier} S. L. Rolston, Ultracold neutral plasmas, Physics
\textbf{1}, 2 (2008).

\bibitem{complex plasma} M. Bonitz, C. Henning, and D. Block, Complex
plasmas: a laboratory for strong correlations, Rep. Prog. Phys. \textbf{73},
066501 (2010).

\bibitem{Rolston} M. Lyon and S. L. Rolston, Ultracold neutral plasmas, Rep.
Prog. Phys. \textbf{80}, 017001 (2017).

\bibitem{Manfredi1} G. Manfredi and F. Haas, Self-consistent fluid model for
a quantum electron gas, Phys. Rev. B \textbf{64}, 075316 (2001).

\bibitem{Manfredi2} G. Manfredi, How to model quantum plasmas, Fields
Institute Communications \textbf{46}, 263-287 (2005).

\bibitem{Shukla} P. K. Shukla and B. Eliasson, Formation and dynamics of
dark solitons and vortices in quantum electron plasmas, Phys. Rev. Lett.
\textbf{96}, 245001 (2006).

\bibitem{Reinisch} G. Reinisch, Nonlinear quantization of a degenerate
charged Bose gas in an external Coulomb trap, Phys. Rev. A \textbf{70},
033613 (2004).

\bibitem{composite} M. Combescot, O. Betbeder-Matibet, and F. Dubin, The
many-body physics of composite bosons, Phys. Rep. \textbf{463}, 215-318
(2008).

\bibitem{Fedele} R. Fedele, F. Tanjia, S. De Nicola, D. Jovanovic, and P. K.
Shukla, Quantum ring solitons and nonlocal effects in plasma wake field
excitations, Phys. Plasmas \textbf{19}, 102106 (2012).

\bibitem{NJP} R. Rugango, J. E. Goeders, T. H. Dixon, J. M. Gray, N. B.
Khanyile, G. Shu, R. J. Clark, and K. R. Brown, Sympathetic cooling of
molecular ion motion to the ground state, New J. Phys. \textbf{17}, 035009
(2015).

\bibitem{Weizmann} Z. Meir, T. Sikorsky, R. Ben-shlomi, N. Akerman, M.
Pinkas, Y. Dallal, and R. Ozeri, Experimental apparatus for overlapping a
ground-state cooled ion with ultracold atoms, J. Mod. Opt. \textbf{65},
501-519 (2018).

\bibitem{Weizmann2} Z. Meir, M. Pinkas, T. Sikorsky, R. Ben-shlomi, N.
Akerman, and R. Ozeri, Direct observation of atom-ion nonequilibrium
sympathetic cooling, Phys. Rev. Lett. \textbf{121}, 053402 (2018).

\bibitem{Holland0} H. A. F\"{u}rst, N. V. Ewald, T. Secker, J. Joger, T.
Feldker, and R. Gerritsma, Prospects of reaching the quantum regime in Li-Yb$%
^{+}$ mixtures, J. Phys. B: At. Mol. Opt. Phys. \textbf{51}, 195001 (2018).

\bibitem{Holland} T. Feldker, H. F\"{u}rst, H. Hirzler, N. V. Ewald, M.
Mazzanti, D. Wiater, M. Tomza, and R. Gerritsma, Buffer gas cooling of a
trapped ion to the quantum regime, Nature Phys.
https://doi.org/10.1038/s41567-019-0772-5 (2020).

\bibitem{Kolomeisky} E. B. Kolomeisky, T. J. Newman, J. P. Straley, and X.
Qi, Low-dimensional Bose liquids: Beyond the Gross-Pitaevskii approximation,
Phys. Rev. Lett. \textbf{85}, 1146-1149 (2000).

\bibitem{Fermi1} S. K. Adhikari, Nonlinear Schr\"{o}dinger equation for a
superfluid Fermi gas in the BCS-BEC crossover, Phys. Rev. A \textbf{77},
045602 (2008).

\bibitem{Fermi2} S. K. Adhikari and L. Salasnich, Superfluid Bose-Fermi
mixture from weak coupling to unitarity, Phys. Rev. A \textbf{78}, 043616
(2008).

\bibitem{Jin} B. De Marco and D. S. Jin, Onset of Fermi degeneracy in a
trapped atomic gas, Science \textbf{285}, 1703-1706 (1999).

\bibitem{diff-mass1} A. Y. Woong, D. L. Mamas, and D. Arnush, Phys. Fluids
18, 1489-1493 (1975).

\bibitem{diff-mass2} J. P. Schermann and F. G. Major, Characteristics of
electron-free plasma confinement in an rf quadrupole field, Appl. Phys.
\textbf{16}, 225-230 (1978).

\bibitem{equal-mass} W. Oohara and R. Hatakeyama, Pair-ion plasma generation
using fullerenes, Phys. Rev. Lett. \textbf{91}, 2-5005 (2003).

\bibitem{deuterium} M. Bacal and M. Wada, Negative hydrogen ion production
mechanisms, Appl. Phys. Rev. \textbf{2}, 021305 (2015).

\bibitem{Ichimaru} S. Ichimaru, Strongly coupled plasmas: high-density
classical plasmas and degenerate electron liquids, Rev. Mod. Phys. \textbf{54%
}, 1017-1059 (1982).

\bibitem{Astra-Girar} G. E. Astrakharchik and M. D. Girardeau, Exact
ground-state properties of a one-dimensional Coulomb gas, Phys. Rev. B
\textbf{83}, 153303 (2011).

\bibitem{positive-negative} F. Robicheaux, B. J. Bender, and M. A. Phillips,
Simulations of an ultracold, neutral plasma with equal mass for every
charge, J. Phys. B: At. Mol. Opt. Phys. \textbf{47}, 245701 (2014).

\bibitem{Pit} M. Kr\"{a}mer, C. Menotti, L. Pitaevskii, and S. Stringari,
Bose-Einstein condensates in 1D optical lattices: Compressibility, Bloch
bands and elementary excitations, Eur. Phys. J. D \textbf{27}, 247-261
(2003).

\bibitem{Smerzi} C Menotti, A Smerzi, and A Trombettoni, Superfluid dynamics
of a Bose-Einstein condensate in a periodic potential, New J. Phys. \textbf{5%
}, 112 (2003).

\bibitem{Carusotto} I. Carusotto and G. C. La Rocca, Modulated optical
lattice as an atomic Fabry-Perot interferometer, Phys. Rev. Lett. \textbf{84}%
, 399-403 (2000).

\bibitem{Inguscio} L. Fallani, F. S. Cataliotti, J. Catani, C. Fort, M.
Modugno, M. Zawada, and M. Inguscio, Optically induced lensing effect on a
Bose-Einstein condensate expanding in a moving lattice, Phys. Rev. Lett.
\textbf{91}, 240405 (2003).

\bibitem{Fermi} M. Iskin and C. A. R. Sa de Melo, Two-species fermion
mixtures with population imbalance, Phys. Rev. Lett. \textbf{97}, 100404
(2006).

\bibitem{mediated} S. K. Adhikari, B. A. Malomed, L. Salasnich, and F.
Toigo, Spontaneous symmetry breaking of Bose-Fermi mixtures in double-well
potentials, Phys. Rev. A \textbf{81}, 053630 (2010).

\bibitem{Dong} J. Qin, G. Dong, and B. A. Malomed, Hybrid
matter-wave-microwave solitons produced by the local-field effect, Phys.
Rev. Lett. \textbf{115}, 023901 (2015).

\bibitem{IT} X. Antoine, W. Bao, and C. Besse, Computational methods for the
dynamics of the nonlinear Schr\"{o}dinger/Gross-Pitaevskii equations, Comp.
Phys. Commun. \textbf{184}, 2621-2633 (2013).

\bibitem{Mineev} V. P. Mineev, Theory of solution of two almost perfect Bose
gases, Zh. Eksp. Teor. Fiz.\textbf{\ 67}, 263-272 (1974) [English
translation: Sov. Phys. JETP \textbf{40}, 132 (1974)].

\bibitem{Salasnich} L. Salasnich, A. Parola, and L. Reatto, Effective wave
equations for the dynamics of cigar-shaped and disk-shaped Bose condensates,
Phys. Rev. A \textbf{65}, 043614 (2002).

\bibitem{Delgado} A. Mu\~{n}oz Mateo and V. Delgado, Effective mean-field
equations for cigar-shaped and disk-shaped Bose-Einstein condensates, Phys.
Rev. A \textbf{77}, 013617 (2008).

\bibitem{Langmuir} H. K. Malik, Oscillating two stream instability of a
plasma wave in a negative ion containing plasma with hot and cold positive
ions, Laser and Particle Beams \textbf{25}, 397-406 (2007).

\bibitem{Agrawal} G. P. Agrawal, \textit{Nonlinear Fiber Optics} (Academic
Press: San Diego, 2007).

\bibitem{House}
J. E. House, \textit{Inorganic Chemistry} (Elsevier, Amsterdam, 2013).


\bibitem{progress} B. A. Malomed, Variational methods in nonlinear fiber
optics and related fields. Prog. Optics 43, 71-193 (2002).

\bibitem{old} B. A. Malomed, A. A. Nepomnyashchy, and M. I. Tribelsky,
Domain boundaries in convection patterns, Phys. Rev. A \textbf{42},
7244-7263 (1990).

\bibitem{Staliunas} K. Staliunas and V. J. S\'{a}nchez-Morcillo, \textit{%
Transverse Patterns in Nonlinear Optical Resonators} (Springer: Berlin,
2003).
\end{thebibliography}
\end{document}